\documentclass[conference,a4paper,10pt]{IEEEtran}
 \usepackage{mathtools} 
 \usepackage{amsfonts}
 \usepackage{amsmath}
 \usepackage{amsthm}
 \usepackage{amssymb}

\usepackage{tablefootnote}
\usepackage{url}

\usepackage{algorithm}
\usepackage{algorithmicx}
\usepackage{algpseudocode}
\usepackage{booktabs}
\usepackage{enumitem}
\usepackage{subfigure}
\usepackage{graphicx}
\usepackage{xcolor}
\usepackage{url}
\usepackage{cite}  
\usepackage{soul}
\usepackage{balance}
\usepackage{multirow}

\ifCLASSINFOpdf
\else
\fi
\begin{document}

%
\title{CSI Measurements and Initial Results for Massive MIMO to UAV Communications}
\author{Zhuangzhuang~Cui\textsuperscript{1}, Achiel Colpaert\textsuperscript{1, 2}, and Sofie~Pollin\textsuperscript{1}\\
\textsuperscript{1}Department of Electrical Engineering (ESAT), KU Leuven, 
\textsuperscript{2}IMEC, Leuven, Belgium\\
Email: \texttt{\{zhuangzhuang.cui, achiel.colpaert, sofie.pollin\}@kuleuven.be}}

%



\maketitle

\begin{abstract}
Non-Terrestrial Network (NTN) has been envisioned as a key component of the sixth-generation (6G) mobile communication system. Meanwhile, unmanned aerial vehicles (UAVs) play an important role in enabling and deploying NTNs. In this paper, we focus on massive multi-input multi-output (MaMIMO) supported UAV communications, where channel state information (CSI) was measured considering different heights and trajectories of a rotary-wing drone. To characterize the propagation channel for this air-to-ground link, some initial results were analyzed, such as stationary distance. To investigate the impact of channels on communication performance, we analyzed spectral efficiency (SE) by using Maximum Ratio Combining (MRC). This study shows that the presented space-time-frequency channel dataset facilitates channel correlation analysis and supports performance evaluation for MaMIMO-UAV communications.
\end{abstract}


\section{Introduction}
Non-Terrestrial Network (NTN) has become one of the key applications in the sixth-generation (6G) mobile communications. As an enabler in the NTN, unmanned aerial vehicles (UAVs) have received much attention, thanks to their flexible deployment, easy manipulation, and low cost \cite{uav_bg}. To consider UAV as an aerial user, how a ground base station (BS) can provide reliable connectivity becomes more promising in the current stage, which is also a feasible solution for beyond line-of-sight (BLOS) connection of UAV users in cellular networks. To this end, the understanding of air-to-ground (AG) channels and the proposal of the corresponding models are the very first steps for the real deployment. However, most existing work in AG channel modeling considers the single-antenna BS, which deviates from the current multi-antenna BS in 5G or 6G. The shortage in massive multi-input multi-output (MaMIMO) channel characterization motivates us to develop the testbed and conduct the measurements.  

In the state of the art, existing work mostly focuses on geometry-based stochastic modeling (GBSM), such as in \cite{jinpeng_uav} where the authors proposed three-dimensional (3D) space-time-frequency non-stationarity modeling for MaMIMO-UAV channels. It is known that GBSM methods can provide a more general methodology of channel realization, by assuming the stochastic distributions of scatterers and the associated channel parameters. However, its effectiveness needs to be further verified by real-world measurements. Regarding measurements considering both MIMO and UAV components, there is very limited work. The reasons lie in the cost, the complicated hardware construction, and over-the-air synchronization while employing a UAV. Nonetheless, the authors in \cite{willink15} started AG channel measurements considering an $8\times2$ MIMO deployment, where temporal channel characteristics and capacity were analyzed. A recent measurement using a $2\times2$ MIMO setting targeted the capacity analysis of AG links, in which both line-of-sight (LOS) and non-LOS (NLOS) channel states were considered \cite{mimo_mea}. It is seen that massive MIMO is rarely used in existing measurements, moreover, a detailed multidimensional channel characterization is limited. 

To fill the gap, we focus on MaMIMO-UAV channel measurements using the MaMIMO testbed. Then, some initial analysis of the channel state information (CSI) dataset is provided. Overall, our contributions can be summarized as follows: 1) we first provided the design and test of MaMIMO-UAV measurement systems, in terms of hardware and software, which lays the foundation of measurement campaigns; 2) we conducted various measurements in a campus environment, including different heights and trajectories of drone, and the obtained dataset is fully open to the research community \cite{0IMQDF_2023}; and 3) we formulated standard data processing procedures, which enable us to obtain various channel parameters such as delay spread, stationary distance, and antenna correlation.

\section{Measurement}
In this section, we first introduce the MaMIMO-UAV measurement system. Then, we provide details of the measurement scenario, UAV trajectories, and dataset structure. 

\subsection{Measurement System}
\begin{figure}[!t]
  \centering
   \subfigure[Base station and drone]{\includegraphics[width=1.7in]{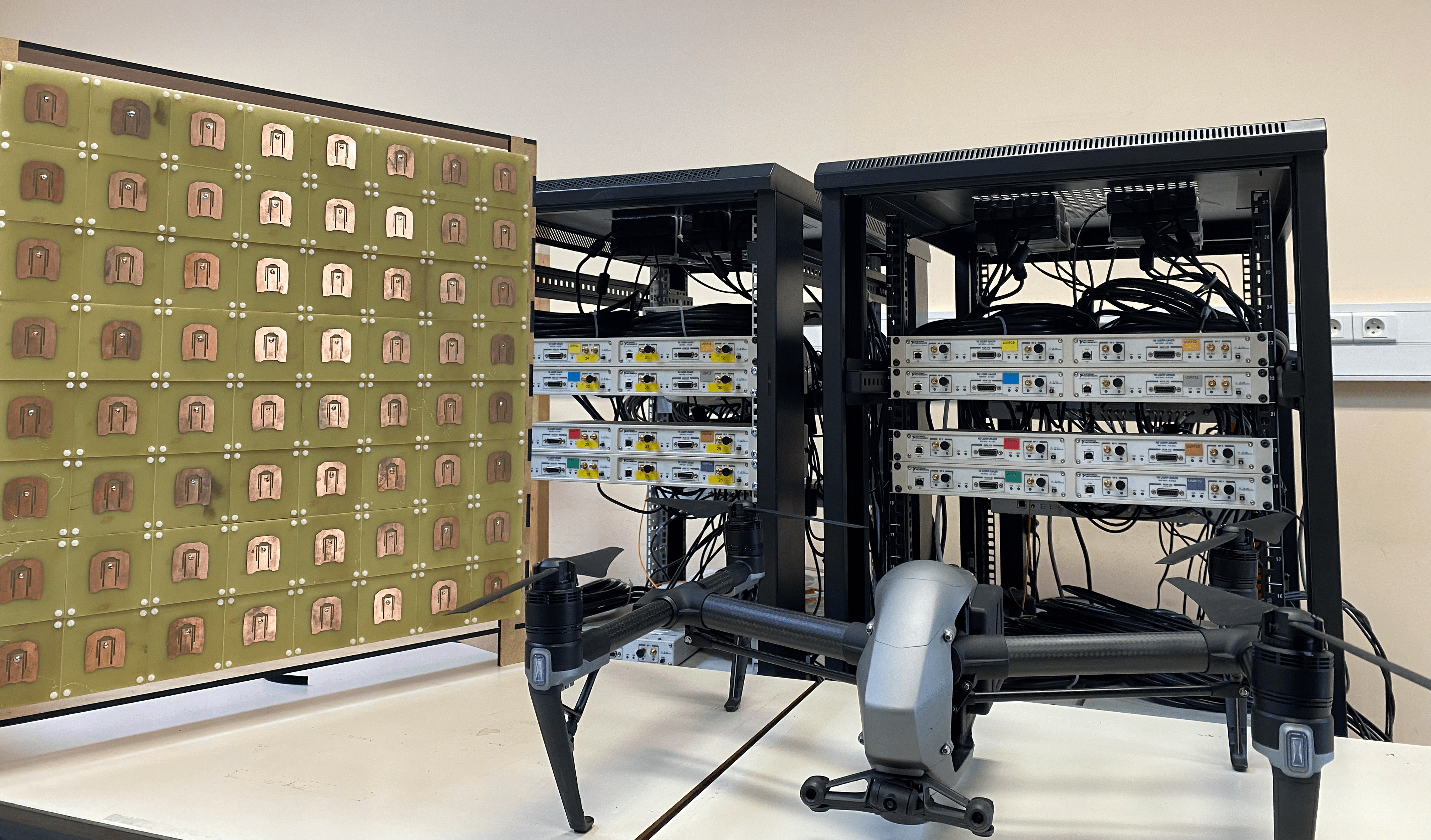}}
   \subfigure[Mobile station ]{\includegraphics[width=1.7in]{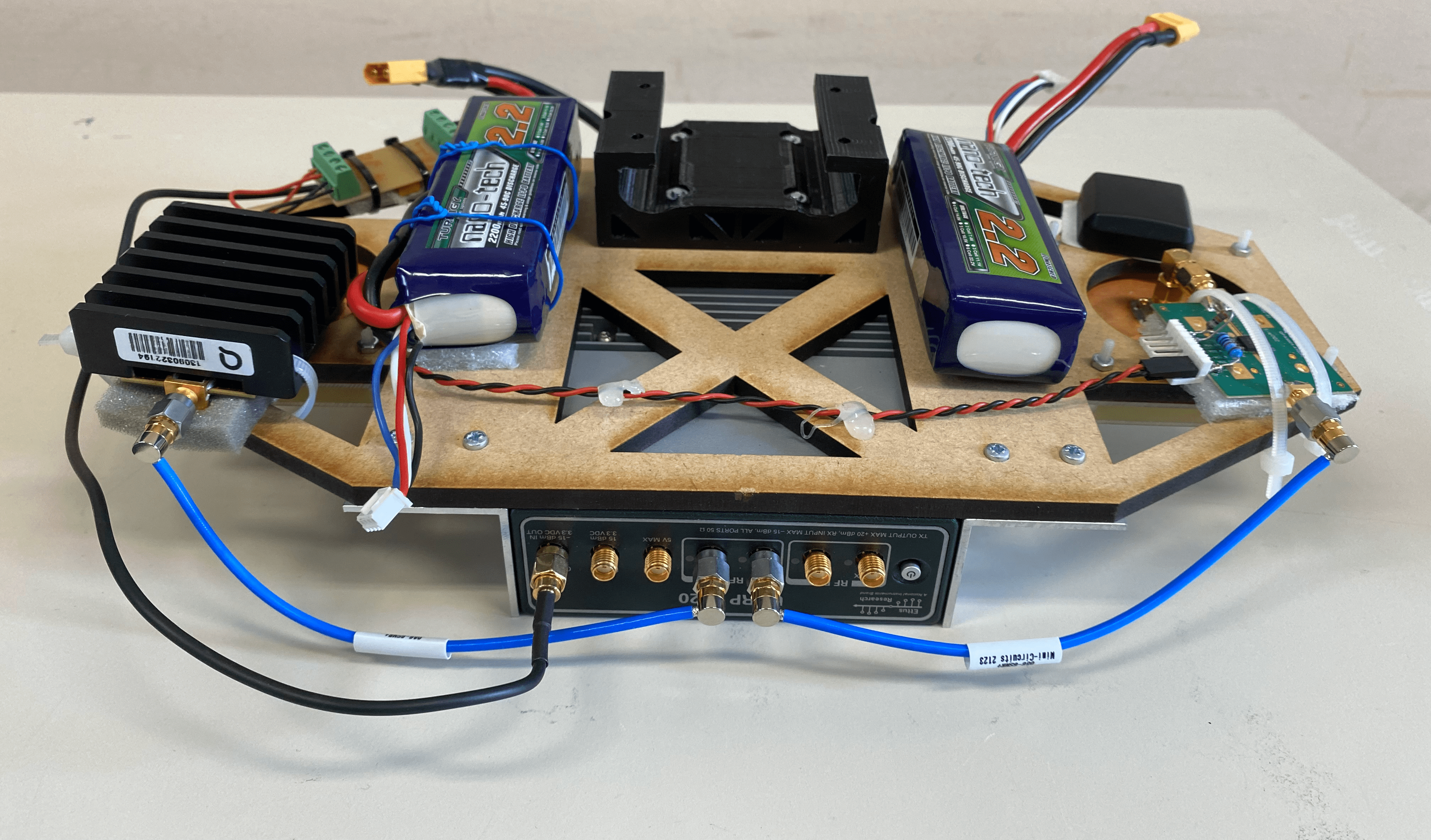}}
  \caption{Measurement system consisting of 32-USRP base station, USRP E320 mobile station, $8\times8$ uniform rectangular array, and a DJI Inspire-2 drone.}
  \label{mamimo-uav}
 \end{figure}

The measurement setup was used in \cite{colpaert20233d}, in which we put the antenna array of the BS along the window and we used patch antennas in the UAV. In this measurement, we place the BS in a parking lot, and two dipole antennas are used in the USRP equipped on the UAV. A BS is equipped with 64 patch antennas specifically designed for an antenna array setup \cite{achiel_mag}. The antennas are arranged in an $8\times8$ Uniform Rectangular Array (URA). The antenna array is 1.2 m above the ground, with its boresight pointed upwards towards the zenith, under a perpendicular angle to the ground. The 64 antennas connect to 32 Software Defined Radios (SDR) that are perfectly synchronized using a shared 10 MHz input reference clock generated by a GPS Disciplined Oscillator. The central frequency is 2.61 GHz with a bandwidth of 18 MHz. The BS uses an LTE-based Time Divison Duplexing frame structure with an Orthogonal Frequency Division Multiplexing (OFDM) signal. All the SDRs collect IQ samples and a central system collects these IQ samples and performs channel estimation every 1~ms. Then, the BS writes these channel estimations to a database file. As shown in Fig.~\ref{mamimo-uav}, we provide pictures of BS, UAV, mobile station, and URA. 

\subsection{Measurement Campaigns}
\begin{figure}[!t]
  \centering
   \includegraphics[width=3in]{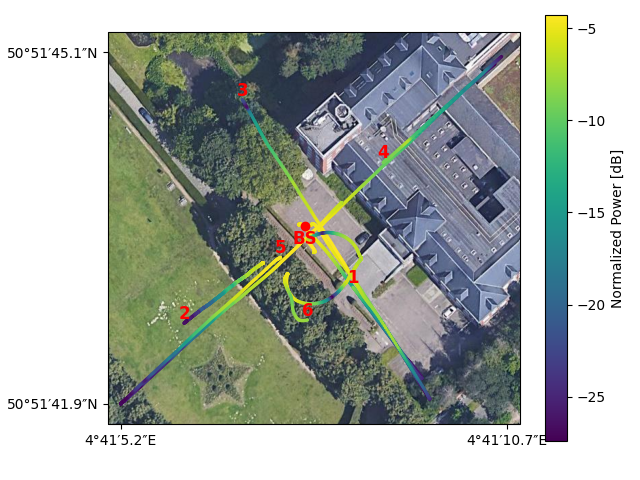}
  \caption{Measurement scenario and trajectories generated by GPS information.}
  \label{environment}
 \end{figure}

\begin{figure}[!t]
  \centering
   \includegraphics[width=2.5in]{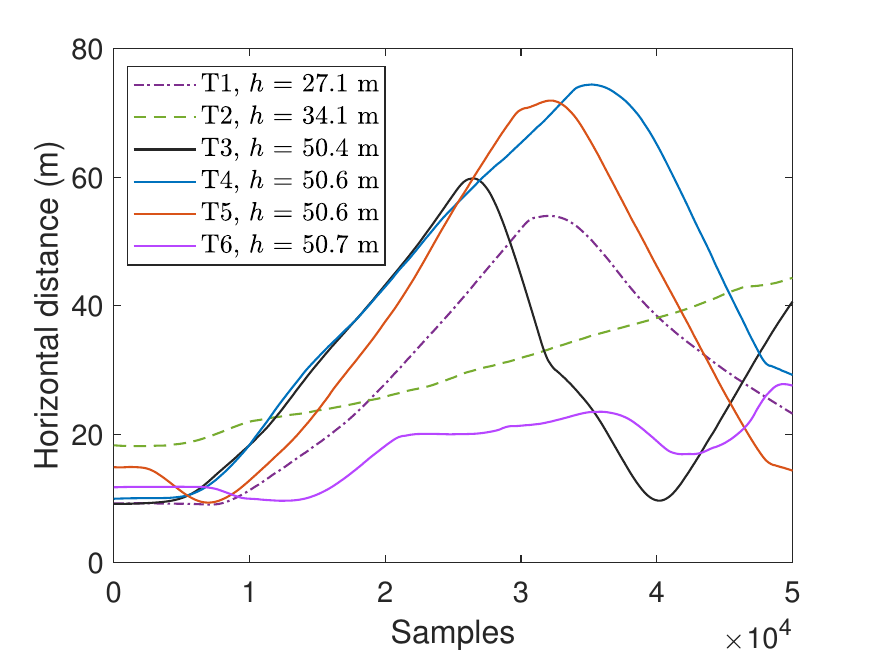}
  \caption{T1-T6 trajectories consisting of the horizontal distances and heights.}
  \label{traj_info}
 \end{figure}

The location of the BS is indicated by the Latitude of $50^{\circ} 51^\prime 43.51917^{\prime \prime}$, the Longitude of $4^{\circ}41^\prime 7.84688^{\prime \prime}$ and the height of 1.2~m. It is noted that the above sea level (ASL) in the measurement environment is around 25~m, which is used to calculate the actual distance between the BS and UAV since the locations of the UAV are recorded by the GPS, in which its height is based on the ASL. In general, six trajectories of the drone were measured, as shown in Fig.~\ref{environment}, where the trajectories are generated by the latitudes and longitudes of GPS information from the drone. Moreover, the normalized power is calculated, which shows the received power decreases with increasing distance. For simplicity, we term these six trajectories as T1-T6, which include different heights and horizontal distances that are plotted in Fig.~\ref{traj_info}. It shows that the altitudes range from 27.1~m to 50.7~m, and the horizontal distances between drone and BS span from 9.1~m to 74.3 m. Moreover, the average speeds of the drone are 1.68~m/s, 0.48~m/s, 3.07~m/s,  2.27~m/s, 2.79~m/s, and 1.36~m/s for T1-T6 trajectory, respectively. The speed remains almost constant during the flight. These flights consist of different propagation conditions including LOS, obstructed-LOS (OLOS), and NLOS. 


\subsection{Data Set Structure}
\begin{figure}[!t]
  \centering
   \includegraphics[width=3.5in]{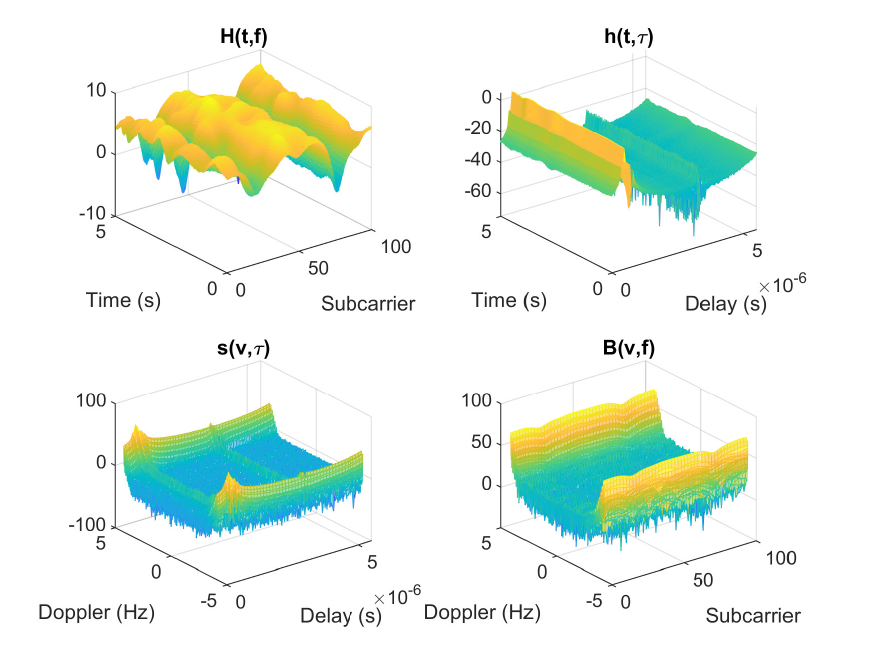}
  \caption{An overview of channel functions in the time-delay-Doppler domain obtained from the measurement data in 0-5 s flight for trajectory T2.}
  \label{channel_functions}
 \end{figure}

For each trajectory, there are 50,000 samples collected with a 1~ms sampling interval, which indicates the flight time is 50~s for one trajectory. The obtained CSI data composes three dimensions, i.e., sample, antenna, and subcarrier, corresponding to the time-space-frequency domain, which enables us to conduct 3D channel analysis. Specifically, the size of raw channel data for each trajectory is $50000\times64\times100$, denoted as $H(t,n,f)$. To visualize the data, we first show the channel transfer function (CTF) and channel impulse response (CIR) in Fig.~\ref{channel_functions}, where the CTF $H(t, f)$ is obtained by averaging 64 antennas, and the CIR $h(t, \tau)$ is obtained by conducting inverse Fourier transform over $f$. Besides, more variants in terms of channel functions can be characterized by the raw data. For instance, we can obtain the Doppler-variant transfer function $B(\nu, f)$ and impulse response $s(\nu, \tau)$ by the Fourier transform over time $t$ of $H(t, f)$ and $h(t, \tau)$. Moreover, the maximum Doppler shift can be calculated by $f_D^{\max}=\frac{v_{\max}}{\lambda}$ where $v_{\max}$ and $\lambda$ are the maximum velocity and wavelength of the central frequency, respectively. Note that Fig.~\ref{channel_functions} shows power profiles, e.g., $20\log_{10}|h(t, \tau)|$, which we term as power delay profile (PDP) in dB, and its linear values calculated by 
\begin{equation}
    P(t, \tau) = |h(t, \tau)|^2. 
\end{equation}

\section{Channel Results}
In this section, we provide the initial results in terms of channel characteristics, including root-mean-square (RMS) delay spread, temporal stationarity, and antenna correlation, which correspond to the frequency, time, and spatial domains. 
\subsection{RMS Delay Spread}
The RMS delay spread is a measure of the multipath effects of wireless channels, which is used for the design of symbol duration to avoid inter-symbol interference (ISI). It is the normalized second-order central moment of the PDP, which can be calculated by \cite{molisch2012wireless}

\begin{equation}
\label{eq:rms}
    S_{\tau}(t_i) = \sqrt{\frac{\sum^{N_{\tau}\Delta\tau}_{\tau=0}P_h(t_i,\tau)\tau^2}{P_m(t_i)}-T_{m}(t_i)^{2}},
\end{equation}
where $i$ is the index of temporal samples, $N_{\tau}$ is the number of delay bin $\Delta\tau$, corresponding to the number of sub-carriers, and the zeroth-order moment, i.e., the time-integrated power $P_m(t_i)$ is calculated by $P_m(t_i)=\sum_{\tau = 0}^{N_{\tau}\Delta\tau}P_h(t_i,\tau)$ with the \textit{mean delay} $T_m(t_i) = \frac{\sum^{N_{\tau}\Delta\tau}_{\tau=0}P_h(t_i, \tau)\tau }{P_m(t_i)}$ \cite{molisch2012wireless}. Moreover, $P_h$ is the averaged PDP over an averaging window $W$ and over $M$ receive antennas (64 elements), defined as follows,
\begin{equation}
    P_h(t_i, \tau) = \frac{1}{MW} \sum_{k = i}^{i+W-1} \sum_{n = 1}^{M} P(t_k, n, \tau),
\end{equation} 
where $P(t_k, n, \tau)$ is the instantaneous PDP, and $W=100$. 
 \begin{figure}[!t]
  \centering
   \includegraphics[width=2.5in]{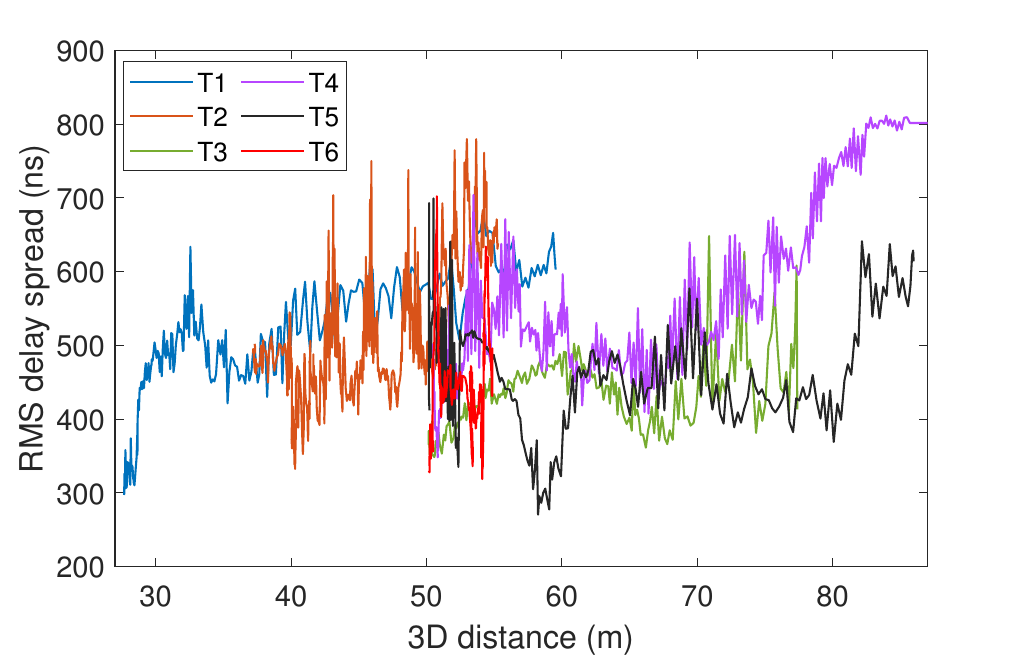}
  \caption{RMS delay spread as a function of 3D distance for T1-T6 trajectories.}
  \label{rms_ds}
 \end{figure}

  \begin{figure}[!t]
  \centering
   \includegraphics[width=2.5in]{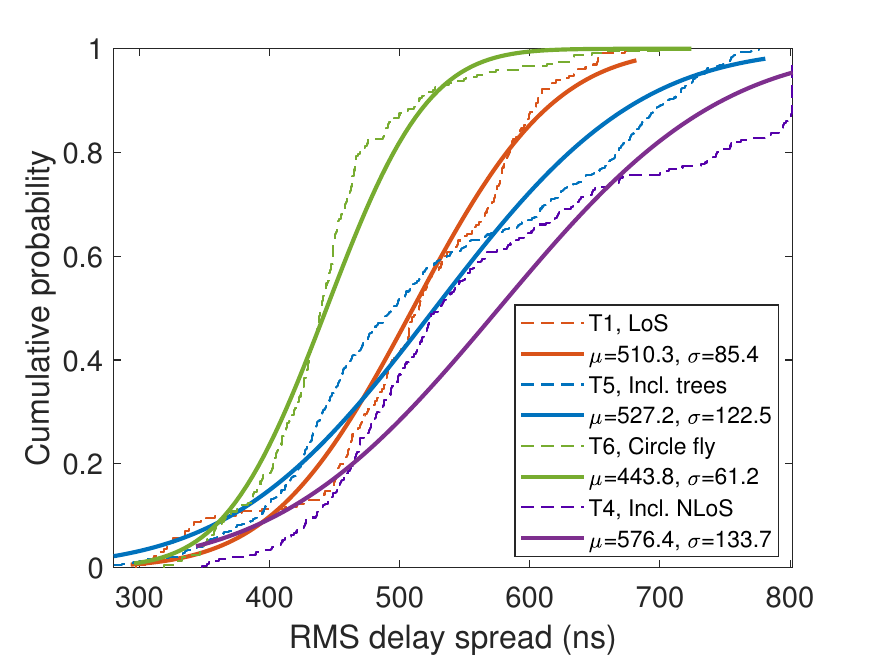}
  \caption{CDF of RMS delay spread for T1, T4, T5, and T6 trajectories.}
  \label{rms_ds_cdf}
 \end{figure}

 For all trajectories, the obtained RMS delay spreads are shown in Fig.~\ref{rms_ds}, which is a function of the 3D distance between BS and drone. On average, the mean values for T1-T6 are around 500~ns, which illustrates that RMS delay spread is mainly determined by the propagation environment. For the same environment, the scatterers are identically distributed, and the corresponding multipath effects are similar. However, for different trajectories, the standard deviations ($\sigma$) of RMS delay spread become distinguishing, which depends on the flight states. For example, the fluctuations of RMS delay spread for T4 and T5 are more severe than those in other trajectories due to NLOS and OLOS conditions. To obtain the statistical results, in Fig.~\ref{rms_ds_cdf}, we provide the cumulative probability distribution function (CDF) of the RMS delay spread for the selected trajectories including T1, and T4-6. It is observed that the circle flight in T6 has the lowest $\sigma$ because of the small changes in 3D distances. As expected, T4 and T5 have high $\sigma$ due to NLOS and OLOS channel states. 
 
\subsection{Temporal Channel Stationarity}
In the time domain, radio channels present a certain similarity in the short time of mobile environment. The stationarity measure is used to characterize this similarity. There are several methods to quantify the stationarity, including temporal PDP correlation coefficient (TPCC) \cite{david_ag} and spectral divergence (SD) \cite{cuieucap}, whose core methodologies are the same, one is for correlation, and the other is for divergence. By using the SD method in the paper, we can first obtain the divergence matrix, expressed as \cite{ruisi} 
\begin{equation}
\label{sd_matrix}
    \gamma (i,j) = \log \left(\frac{1}{N_{\tau}^2}\sum_{p=1}^{N_{\tau}}\frac{P_h(i t,p\tau)}{P_h(j t,p\tau)}\sum_{p=1}^{N_{\tau}}\frac{P_h(j t,p\tau)}{P_h(i t,p\tau)}\right).
\end{equation}
Then, the stationary distance is defined by the region over which the SD stays below a certain threshold, $c_{th}$, defined by
\begin{equation}
\label{eq:sd}
    d_{SD} = v(t_{max} - t_{min}),
\end{equation}
where $v$ is the average speed of drone, and $t_{max}$ and $t_{min}$ are the bounds of the stationarity interval, given by
\begin{equation}
\label{eq:sd_int}
\begin{split}
    & t_{min} = \operatorname*{arg\,max}_{0 \leq j \leq i-1} \gamma (t_i,t_j) \leq c_{th}, \\
    & t_{max} = \operatorname*{arg\,min}_{i+1 \leq j \leq T - W} \gamma (t_i,t_j) \leq c_{th}. \\
\end{split}
\end{equation}

  \begin{figure}[!t]
  \centering
   \includegraphics[width=2.5in]{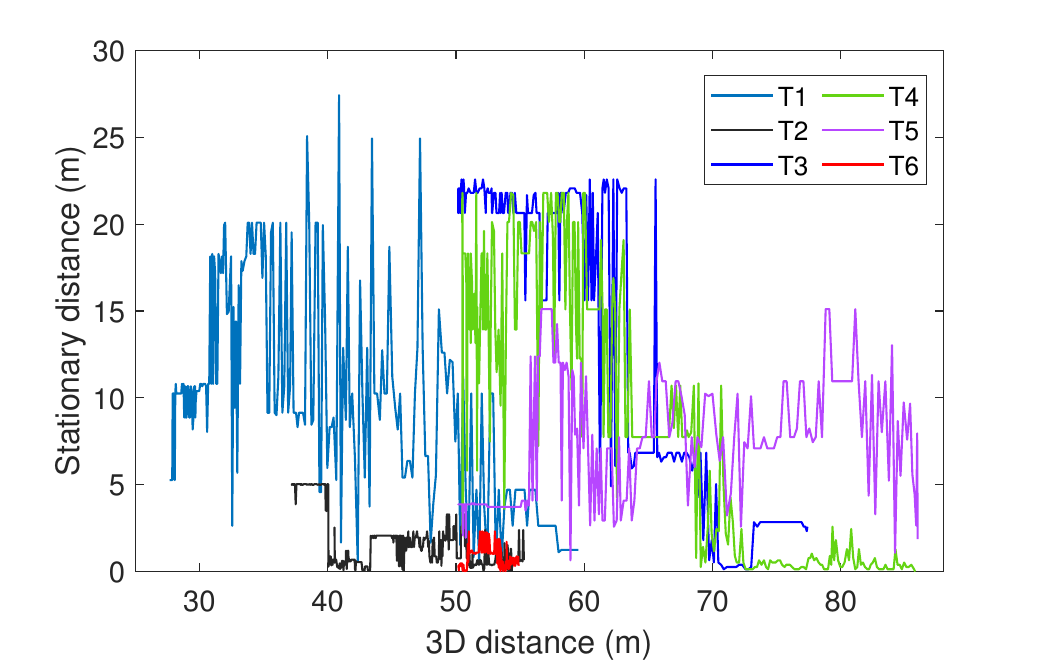}
  \caption{Stationary distance versus 3D drone-BS distance for all trajectories.}
  \label{sd_traj}
 \end{figure}
\begin{figure}[!t]
  \centering
   \includegraphics[width=2.5in]{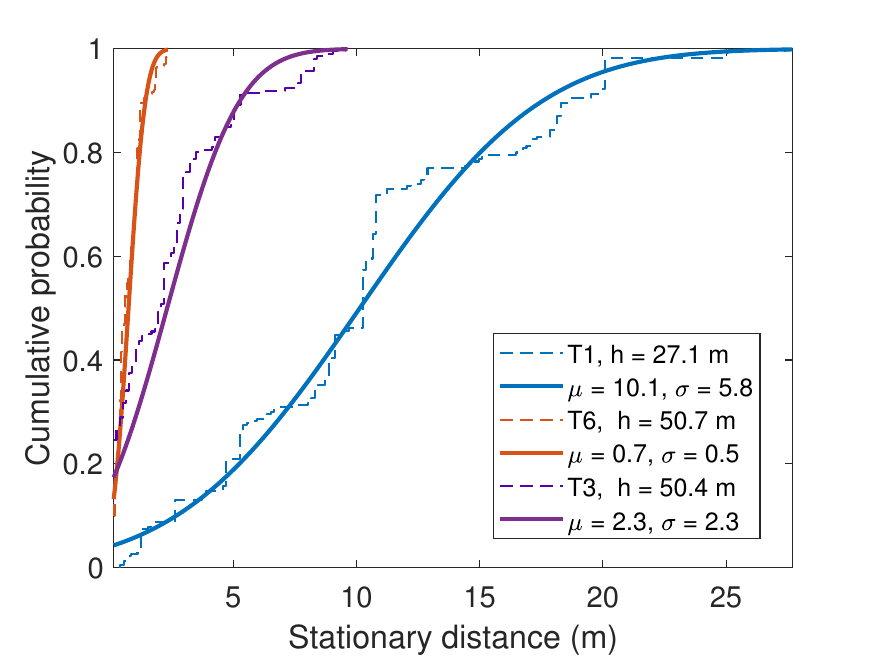}
  \caption{CDF of stationary distance for T1, T3, and T6 trajectories.}
  \label{sd_cdf}
 \end{figure}

Considering the threshold $c_{th}=e^{-1}$, we obtain the stationary distance results which are shown in Fig.~\ref{sd_traj}, where the 3D distance of the $x$-axis is as a function of the flight time. It is found that the stationary distances are more than 15 meters in the most of LOS conditions of trajectories, such as T1, T3, and T4. For NLOS and OLOS conditions, such as T5 and part of T4, the stationary distances become smaller, which indicates the channel non-stationarity is more severe. For T6, the stationary distance is short because the total traveled distance in this flight is less than 10 meters, while the maximum stationary distance is bounded within the total traveled distance. For a fair comparison, we used the normalized stationarity defined by $\frac{d_{SD}}{d_{Traveled}}$ in our previous paper \cite{colpaert20233d}. Besides, it is observed that $d_{SD}$ follows the Gaussian distribution in Fig.~\ref{sd_cdf}, where we select the typical trajectories considering different heights. Both mean $\mu$ and standard deviation ($\sigma$) become smaller when the height changes from T1 to T3, which shows the channel becomes more \textit{clean} while the stationarity also becomes lower.

 \subsection{Antenna Correlation}
 Antenna spatial correlation represents the correlation between elements, in which both channel transfer function or impulse response can be used to calculate the cross-correlation. Therefore, we obtain the correlation between the $n$th and $m$-th element, expressed as \cite{tao16} 
    \begin{equation}
        R(t_i, m,n) =\frac{\mathbb{E}[h_m(t_i) h_n^{*}(t_i)]}{\sqrt{\mathbb{V}[h_m(t_i)]\mathbb{V}[h_n(t_i)]}},
    \end{equation}
where $\mathbb{E}$ and $\mathbb{V}$ represent the expectation and variance respectively. $(\cdot)^*$ indicates the complex conjugation. 

\begin{figure}[!t]
  \centering
   \subfigure[A single element ($8\times8$)]{\includegraphics[width=2in]{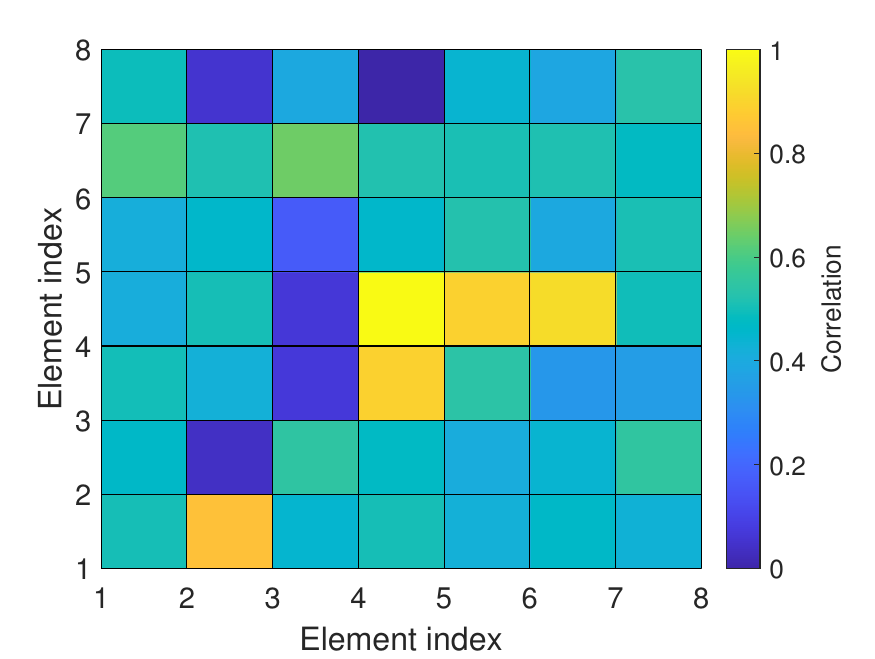}}
   \subfigure[All elements ($64\times64$)]{\includegraphics[width=2in]{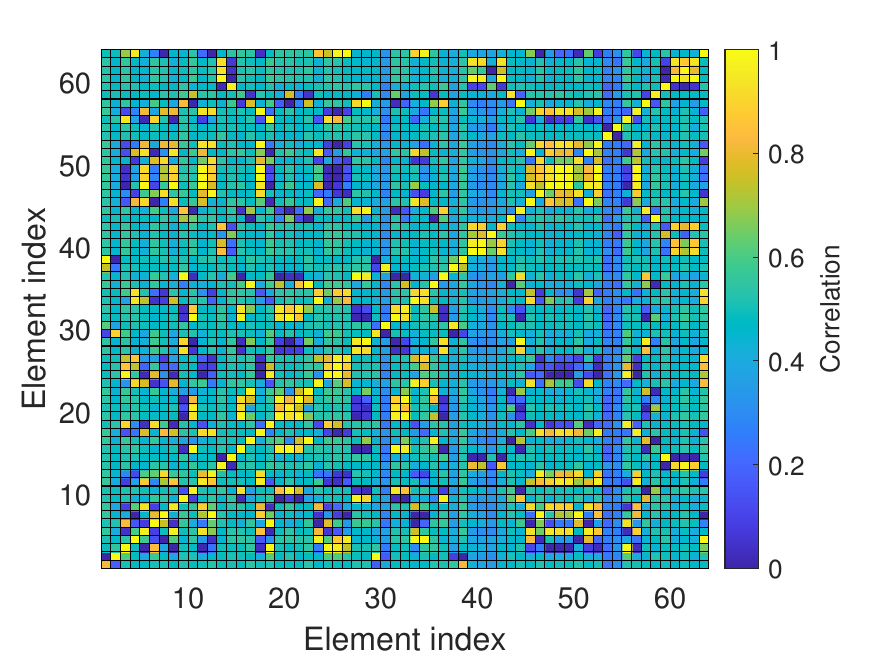}}
  \caption{Antenna correlation for a single element and all elements in T3.}
  \label{antenna}
 \end{figure}

As shown in Fig.~\ref{antenna}(a), the correlation between the element index $(4, 4)$ and all 64 elements is illustrated. It is observed that a high correlation happens with its surrounding elements, which indicates spatial stationarity in an antenna array. We then show the case with all elements in Fig.~\ref{antenna}(b), where the diagonals indicate auto-correlations of the elements themselves. 

A traditional spatial correlation model can be expressed as $J_0^2(2\pi d_s/\lambda)$ where $J_0$ is the Bessel functions of the first kind \cite{molisch2012wireless}. Moreover, $d_s$ can be the displacement of the transceiver or the spacing of elements in an array. It shows this model agrees with our measurement results, which also corresponds to \textit{Jakes} (U-shaped) spectrum of $s(\nu, \tau)$ in Fig.~\ref{channel_functions}, assuming a uniform angular spectrum in azimuth. 

\section{Performance and Mobility}
In this section, the CSI is used to obtain spectral efficiency, and trajectory data is used to discuss the wobbling effects.
\subsection{Spectral Efficiency}
\begin{figure}[!t]
  \centering
   \includegraphics[width=2.5in]{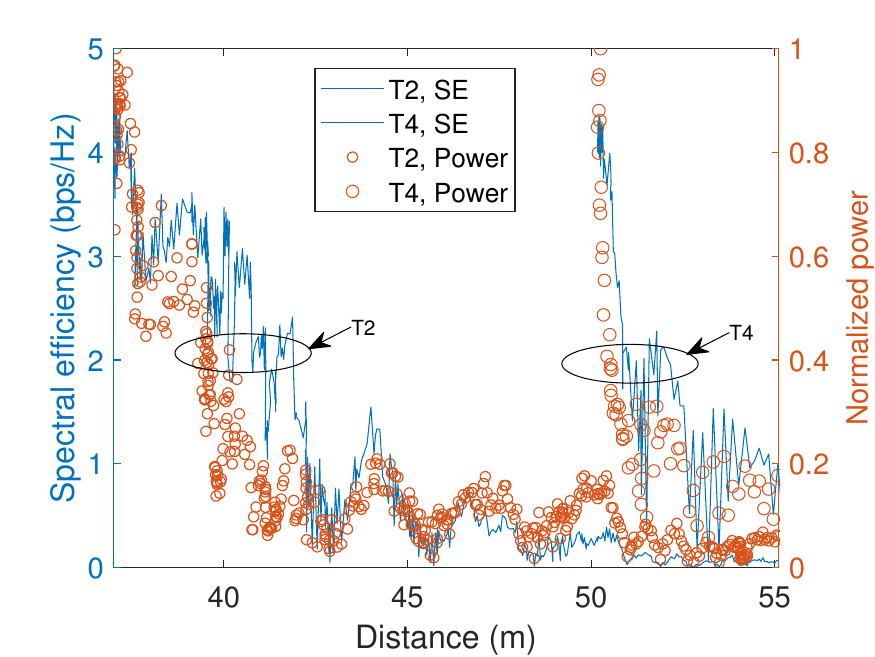}
  \caption{Spectral efficiency versus 3D distance for T2 and T4 trajectories.}
  \label{se_t24}
 \end{figure}

We calculate the spectral efficiency based on its definition, and the maximum ratio combining (MRC) is used to combine all antenna elements ($M=64$) with weights $w_n=h_n^{*}$ for $n=1...M$. We show the results for the selected T2 and T4 trajectories in Fig.~\ref{se_t24}. Moreover, for each snapshot, we include the maximum normalized power as a comparison. It shows both SE and received power have the same trend. Moreover, the SE deterioration is slower in T2 when the channel changes from LOS to OLOS, compared to that in T4 with the transition from LOS to NLOS. One can conclude that it is more important to guarantee the LOS connectivity, and less important to control the flight distance between UAV and BS in practical drone deployment.
   
\subsection{Mobility}
\begin{figure}[!t]
  \centering
   \includegraphics[width=2.5in]{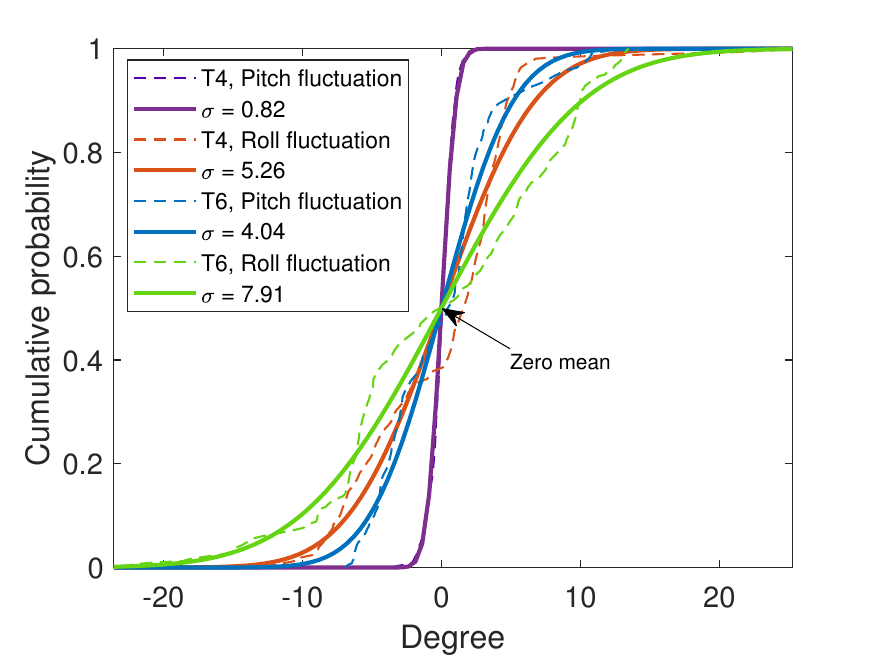}
  \caption{CDF of pitch and roll fluctuations for T4 and T6 trajectories.}
  \label{pitch_roll}
 \end{figure}

The six-dimensional (6D) mobility including translation and rotation has been of interest in UAV channel modeling \cite{molisch20}. In the trajectory data, we can extract the pitch and roll angles of the drone flights. We select two typical trajectories straight-lined T4 and circled T6 for the discussion, where the fluctuations of pitch and roll angles are indicated by their standard deviations. In Fig.~\ref{pitch_roll}, the CDFs of these fluctuations are given. It is found the pitch has a more slight wobbling effect than that in terms of roll. Besides, the circle trajectory leads to high fluctuations for both pitch and roll. Nonetheless, an observation suggests that all these fluctuations follow a zero-mean Gaussian distribution, which provides experimental evidence for the theoretical analysis work. 
 
\section{Summary and Conclusion}
\begin{figure}[!t]
  \centering
   \includegraphics[width=2.7in]{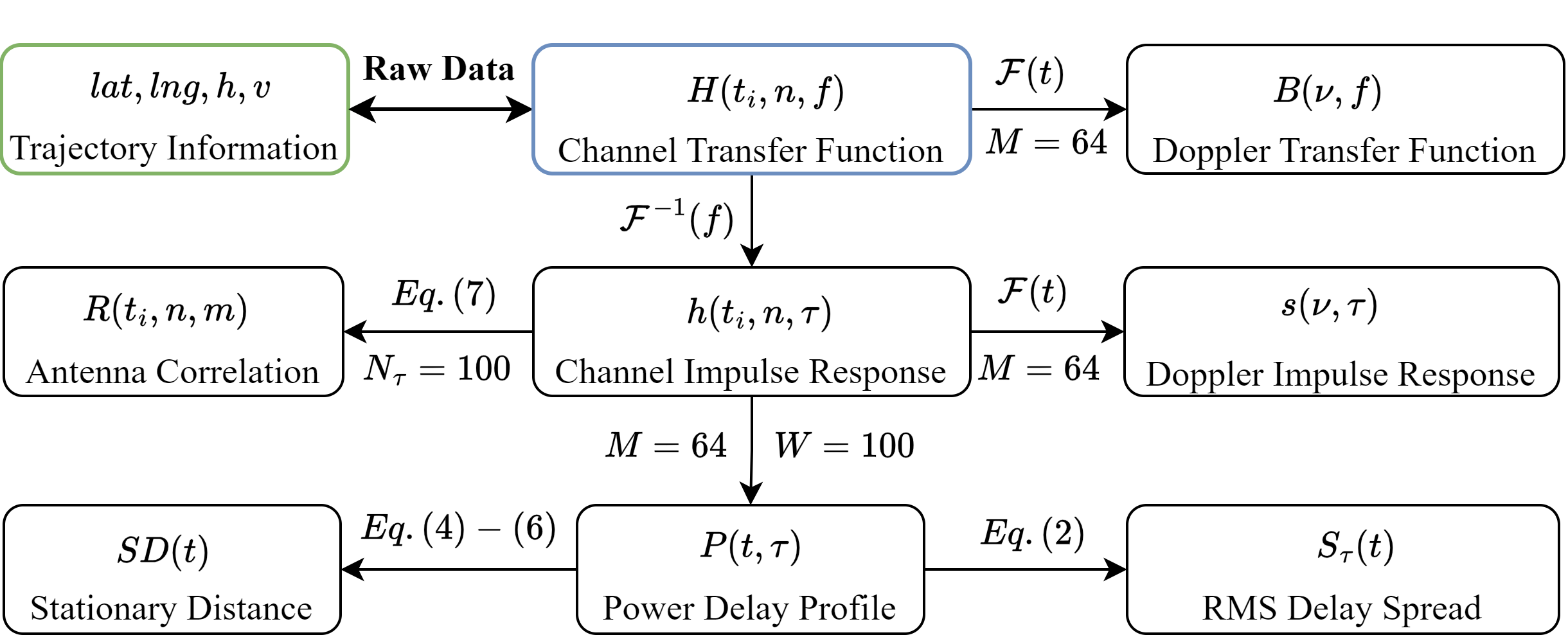}
  \caption{Summary of procedures of channel data processing in this paper.}
  \label{channel_pro}
 \end{figure}

This paper presents MaMIMO-UAV channel measurement campaigns and corresponding channel data processing. As a summary, we include the main procedures in Fig.~\ref{channel_pro}, in which the raw data consists of the trajectory data and channel information in the space-time-frequency domain. With (inverse) Fourier transform over different dimensions, we can obtain various channel functions. Then, we analyzed the frequency dispersion indicated by the RMS delay spread, the temporal stationarity, and the spatial correlation in the antenna array. 

Through initial analysis of CSI data for different trajectories, we found RMS delay spreads are around 500~ns, which is mainly environmentally determined. However, the unfavorable channel states (OLOS and NLOS) lead to an increase, which can be up to 800~ns in T4. A similar phenomenon happens in the stationary distance, which indicates the LOS state has higher values. Besides, the SE performance also shows high sensitivity to the channel state, rather than the communication distance. Finally, the impact of 6D mobility considering the wobbling effects of UAV can be considered in our future work.

\section*{Acknowledgement}
This work is supported by the Research Foundation Flanders (FWO), project No. G098020N, and by the iSEE-6G project under the Horizon Europe Research and Innovation programme with Grant Agreement No. 101139291.

\bibliographystyle{IEEEtran}
\bibliography{Reference}


\end{document}